\documentclass[11pt,twoside]{article} 
\usepackage{intlp}
\copyrightnotice{2002}{6}{307}{327}	

\usepackage{amssymb,amsthm}             
\usepackage[mathscr]{eucal}		

\numberwithin{equation}{section}        

\newcommand{\mnote}[1]{}  		

\renewcommand{\Re}{{\mathbb R}}         
\newcommand{\half}{\frac{1}{2}}         

\newcommand{\tr}{\text{\rm tr}}		
\newcommand{\Lie}{\mathcal L}
\renewcommand{\a}{\alpha}
\renewcommand{\b}{\beta}

\newcommand\beq{\begin{eqnarray}}
\newcommand\eeq{\end{eqnarray}}
\newcommand\ben{\begin{enumerate}}
\newcommand\een{\end{enumerate}}

\newcommand{\scri}{\mathscr I}
\newcommand{\onScri}{\big{|}_{\scri}}
\newcommand{\onNt}{\big{|}_{N_t}}
\newcommand{\CC}{\mathcal C}

\newcommand{\tg}{\tilde g}
\newcommand{\tD}{\tilde D}
\newcommand{\tR}{\tilde R}
\newcommand{\tK}{\tilde K}
\newcommand{\tH}{\tilde H}

\newcommand{\tM}{\tilde M}
\renewcommand{\th}{\tilde h}

\newcommand{\bOmega}{\bar \Omega}
\newcommand{\bg}{\bar g}
\newcommand{\br}{\bar r}

\newcommand{\Ric}{\text{\rm Ric}}

\theoremstyle{plain}
\newtheorem{thm}{Theorem}[section]

\newtheorem{cor}[thm]{Corollary}

\newtheorem{prop}[thm]{Proposition}
\theoremstyle{remark}
\newtheorem{remark}{Remark}[section]

\allowdisplaybreaks[2]

\begin{document}


\title{dS/CFT and spacetime topology}
\url{hep-th/0202161}
\author{Lars Andersson$^{a\dagger}$, Gregory J. Galloway$^{b\star}$}
\renewcommand{\thefootnote}{}
\footnotetext{$^\dagger$Supported in part by the Swedish Natural
Sciences Research Council (SNSRC),  contract no.  R-RA 4873-307 and NSF,
contract no. DMS 0104402.}
\footnotetext{$^\star$Supported in part by the NSF,
contract no. DMS 0104042.}

\address{Department of Mathematics\\
University of Miami\\
Coral Gables, FL 33124\\
USA}
\addressemail{$^a$ larsa\char'100math.miami.edu, $^b$galloway\char'100math.miami.edu}

\markboth{\it dS/CFT and spacetime topology}{\it L.~Andersson and G.~J.~Galloway}



\section{Introduction}
In the general Euclidean formulation of the AdS/CFT correspondence put
forth in \cite{witten:AdS-holog},
one considers Riemannian manifolds of the form $M^{n+1}\times Y$, where
$M^{n+1}$ is conformally
compactifiable with conformal boundary-at-infinity $N^n$.
It is then a problem of fundamental interest to determine, for a given
compact manifold $N$
with a given conformal structure, the complete Einstein manifolds of
negative Ricci curvature
having $N$ as conformal boundary.  An early result of this type, predating
AdS/CFT,
was obtained by Graham and Lee \cite{graham:lee}, who showed that for a
conformal structure
sufficiently close to the standard one on $S^n$, there is a unique Einstein
metric with a
prescribed curvature close to the standard hyperbolic metric on the $n+1$
ball which induces the
conformal boundary $N$.
\cutpage

More recently, motivated by certain issues in AdS/CFT,
Witten and Yau~\cite{witten:yau} obtained some general topological
restrictions on
$M$.
They proved in this context that if $M$ is an Einstein manifold with negative
Ricci curvature,
such that  the conformal class of $N$ admits a metric
of positive scalar curvature then the co-dimension one homology of $M$
vanishes,
$H_n(M,\mathbb Z) = 0$. This implies,
in particular, that $N$ is connected.  As discussed in
\cite{witten:yau,witten:talk},
these results resolve certain ``puzzles" concerning the AdS/CFT correspondence.
These, and related results, were
extended by Cai and Galloway
\cite{cai:galloway} to the case of zero scalar curvature;
see also \cite{Wang} for further developments.  This,
in a sense,
covers all cases
relevant to the AdS/CFT correspondence, since,
as argued in \cite{witten:yau}, CFT's defined on
conformal boundaries of negative scalar curvature are unstable.
Results of a related nature in the Lorentzian context for asymptotically
AdS spacetimes
follow from results on topological censorship \cite{GSWW,GSWW2,witten:talk}.

In the present paper we study similar issues
for spacetimes of de Sitter type, i.e.,
spacetimes satisfying the Einstein
equation with cosmological constant $\Lambda >0$, which admit a regular
conformal (Penrose)
compactification.  More specifically, we study the influence of the
curvature and topology
of the conformal boundary (at past or future infinity) on the bulk
spacetime, for
spacetimes of de Sitter type.  Our motivation for this study comes from
recent proposals
for a de Sitter analogue of the AdS/CFT correspondence (see e.g.,
\cite{strominger:ds/CFT:corr,strominger:ds/CFT:infl}), and also from
current developments in cosmology, in
particular the supernova observations which have led cosmologists to
include a positive
$\Lambda$ among the cosmological parameters of the standard model of the
universe, see \cite{perlmutter}.

Consider a spacetime $M$  of de Sitter type  which admits a conformal
completion to the past and future, such that the past conformal boundary
$\scri^-$ and future
conformal boundary $\scri^+$ are spacelike and compact.  When $M$ is
globally hyperbolic,
this implies that $\scri^-$ and $\scri^+$ are each connected, and
homeomorphic, irrespective of
any field equations.  A spacetime $M$ with these properties is nonsingular, in the sense
of being timelike and null geodesically complete.
Although conformal infinity still consists of two components, $\scri^{\pm}$,
there are reasons to view this situation as analogous to a
conformally compactifiable Riemannian  manifold $M$ with \emph{connected}
conformal boundary
$N$, as considered above.  For example, in the dS/CFT  proposal recently
put forward by Strominger
\cite{strominger:ds/CFT:corr}, it is argued that $\scri^+$ and $\scri^-$
become effectively identified,
and give rise to a single conformal field theory; see also \cite{witten:QG}.
At the purely classical level, a recent result of Anderson
\cite{anderson:prescribed}
shows that
the boundary map from
Riemannian AdS metrics on $B^4$, restricted to the connected component
containing the hyperbolic metric, to the component of the
space $\CC^0$ of conformal classes on $S^3$
with positive scalar curvature, containing the round sphere,
has degree one, and is hence surjective
\cite[Theorem C]{anderson:prescribed}.
In the Lorentzian case, results by Friedrich
\cite{friedrich:positive,friedrich:EYM},
suggest a similar relation between $\CC^0 \times \CC^0$ and
the space of asymptotically de Sitter spacetimes on $S^3\times [0,1]$.
Roughly stated, there is an analogy between ``filling in $S^3$" in the
Riemannian AdS
case, and ``filling in two copies of $S^3$" in the Lorentzian de Sitter case.

Thus, for globally hyperbolic spacetimes of de Sitter type with compact
conformal
boundaries $\scri^{\pm}$, the notion of ``connectedness of the boundary" is
in some sense built in.  Our main results then imply that for such spacetimes,
which obey suitable energy conditions, the curvature and topology of
$\scri^+$ and $\scri^-$ are quite restricted: Each must have finite fundamental
group and the associated conformal class of each must contain a metric of
positive
scalar curvature.  
Thus, in analogy with the results of Witten and Yau~\cite{witten:yau}
pertaining to the AdS/CFT correspondence,
we establish here, for asymptotically de Sitter spacetimes, some connections between 
the bulk spacetime (e.g., its being nonsingular)
and the topology of the conformal boundary.  Further discussion of the role of topology
in the dS/CFT correspondence may be found in \cite{mcinnes}.

In the following subsection we give a somewhat more detailed
description of our main results.
From a rather different point of view, our results can be interpreted as
statements
about the topology and completeness of inflationary cosmological models;
see the comment
at the end of the next subsection.

\subsection{Overview of the paper}
We consider spacetimes which are asymptotically de Sitter either to the
future or the past.  To fix the time orientation for the present discussion,
let $M$ be a globally hyperbolic spacetime of de Sitter type with
regular \emph{past} conformal boundary $\scri^-$, see Section
\ref{sec:prelim} for
definitions, and assume $\scri^-$ is compact.  Then the Cauchy surfaces of
$M$ are compact, and in fact
homeomorphic to $\scri^-$, see Proposition \ref{prop:structure}.
Subject to appropriate energy conditions, our results
show that, due to the development of singularities, or  other
irregularities, $M$ cannot
be asymptotically de Sitter to the future; i.e., cannot have a regular
future conformal infinity $\scri^+$,
unless the curvature and topology of $\scri^-$ is suitably restricted.
We briefly discuss here the various curvature and topology
restrictions obtained.

The Riemannian metric
$\th_{\alpha\beta}$ induced by $\tg_{\alpha\beta}$ on $\scri^-$ changes by a
conformal factor
with a change in  the \emph{defining function} $\Omega$, and thus
$\scri^-$ is endowed with a natural conformal structure
$[\th_{\alpha\beta}]$.  By  the
Yamabe theorem, the conformal class $[\th_{\alpha\beta}]$ contains a metric
of constant scalar curvature $-1$,
$0$, or $+1$, exclusively, in which case we will simply say that $\scri^-$
has negative, zero, or positive scalar
curvature, respectively.

In Section \ref{sec:past}, we show that, with the setting as above, if
$\scri^-$ has
negative scalar curvature then
all the timelike geodesics of $M$ are future incomplete.
We further show that if $\scri^-$ has zero scalar curvature,
$M$ can contain a future complete timelike geodesic
only under special circumstances: $M$ must split
as a warped product.
As discussed in Section \ref{sec:past},
these results can be expressed in terms of the
\emph{Yamabe type} of $\scri^-$,
and hence the Yamabe type of the Cauchy surfaces of $M$.
The upshot is, in order for
$M$ to be timelike geodesically complete,
$\scri^-$ must be of positive
Yamabe type.  Hence, in $3+1$ dimensions, $\scri^-$ cannot have any
$K(\pi,1)$ factors
in its prime decomposition.  Thus, modulo the Poincar\'e conjecture,
$\scri^-$ must
be covered by a $3$-sphere, be diffeomorphic to $S^1\times S^2$, or be a
connected
sum of such manifolds.

Some results of a related nature are obtained in Section \ref{sec:simp}.
Corollary \ref{cor:null} shows that with the setting as above,
$M$ cannot admit a regular future conformal boundary $\scri^+$, compact
or otherwise, unless $\scri^-$ has finite fundamental group.  Perhaps
somewhat surprisingly, this rules out, in particular, a scenario
in which a black hole forms from a regular past (with compact~$\scri^-$),
such that $M$, with Cauchy surface topology $S^1\times S^2$, is future
asymptotically similar
to Schwarzschild-de Sitter spacetime.
In a somewhat related vein,
it is shown in Theorem \ref{thm:worm}
that $M^{n+1}$, $n\le 7$, must be future null geodesically incomplete,
unless  $H_{n-1}(\scri^-,\mathbb Z)$ vanishes, or, equivalently, by Poincar\'e
duality,
etc., unless  $H_1(\scri^-,\mathbb Z)$ is pure torsion and finite.
In $3+1$ dimensions, Corollary \ref{cor:null} implies that in order for
$M$ to have a conformal structure similar to that of de Sitter space, $\scri^-$
must be covered by a homotopy $3$-sphere; see also Theorem \ref{thm:null}.
With regard to energy conditions (see Section \ref{sec:prelim}), the results
of Section~\ref{sec:simp} only require the null energy condition.

Finally, we remark that Theorems \ref{thm:incomplete}, \ref{thm:split}, and
\ref{thm:worm},
discussed here in a time dual  manner, can be interpreted as singularity
results, which establish,
as a consequence of certain curvature or topology assumptions,
the occurence of \emph{past} singularities in inflationary
cosmological models.

\section{Preliminaries}\label{sec:prelim}
Let $(M,g_{\alpha\beta})$ be an $n+1$ dimensional
\footnote{Greek indices $\alpha,\beta,\dots$ run
over $0,\dots,n$ while lower case latin indices
$a,b,c,\dots$ run over $1,2,\dots n$.}
space--time, $n \ge 2$, with covariant
derivative $D_{\alpha}$, Ricci tensor $R_{\alpha\beta}$ and scalar curvature
$R$. The Einstein equation with
cosmological constant $\Lambda$ is
\begin{equation}\label{eq:einst}
R_{\alpha\beta} - \half R g_{\alpha\beta} + \Lambda g_{\alpha\beta} =
S_{\alpha\beta},
\end{equation}
where $S_{\alpha\beta}$ is the stress energy tensor.
Let $S = g^{\alpha\beta} S_{\alpha\beta}$.
The stress energy tensor
satisfies the weak, dominant and strong energy
conditions respectively, if for any causal vector field $V^{\alpha}$,
it holds that
\begin{subequations}\label{eq:energycond}
\begin{align}
\text{(W.E.C.)} &&
S_{\alpha\beta} V^{\alpha} V^{\beta} &\geq 0 \label{eq:weakenergy}\\
\text{(D.E.C.)} &&
S_{\alpha\beta} V^{\alpha} V^{\beta} &\geq 0, \quad \text{and} \quad
S^{\alpha\beta}V_{\beta} \text{ is causal } \label{eq:dominant} \\
\text{(S.E.C.)} &&
( S_{\alpha\beta} - \frac{1}{n-1} S g_{\alpha\beta}  ) V^{\alpha}
V^{\beta} &\geq 0  \label{eq:strongenergy}
\end{align}
\end{subequations}
We will consider the case $\Lambda > 0$. After a rescaling, we may
assume that $\Lambda = n(n-1)/2$. With this normalization,
the strong energy condition
(\ref{eq:strongenergy}) is equivalent to
\begin{equation}\label{eq:strong-alt}
R_{\alpha\beta}V^{\alpha}V^{\beta} \geq n g_{\alpha\beta} V^{\alpha}V^{\beta}
\end{equation}
for any causal vector field $V^{\alpha}$. Both the weak and the strong
energy conditions imply the null energy condition
\begin{equation}\label{eq:NEC}
R_{\alpha\beta} X^{\alpha}X^{\beta} \geq 0,
\end{equation}
for any null vector $X^{\alpha}$.

Our proofs occasionally make use of notions
from causal theory.  We briefly recall here
some basic notation, terminology and results; for further
details,  see e.g. \cite{HE,ON}.   For
a subset $A$ of a spacetime $M$, 
the timelike future of $A$, denoted $I^+(A)$, consists
of all points in $M$ that can be reached from $A$ by 
future directed timelike curves.  (Sometimes this is
written as $I^+(A,M)$ to emphasize the 
particular spacetime involved.)  Similarly the 
causal future of $A$, denoted $J^+(A)$, consists of the points of $A$
together with the points in $M$ that can reached from $A$ by future directed causal
curves.  Sets of the form $\partial I^+(A)$ are called achronal boundaries,
and, when nonempty, are  achronal (meaning that no two points can be joined by a timelike
curve) $C^0$ hypersurfaces.
For a closed achronal set $S\subset M$, the future domain of dependence
of $S$, denoted $D^+(S)$,
consists of all points  $p\in M$ such that each past inextendible
causal curve from $p$ meets $S$.  The future Cauchy horizon of $S$, denoted
$H^+(S)$, is the future boundary of $D^+(S)$; one has 
$\partial D^+(S) =  H^+(S) \cup S$.
The sets $I^-(A)$, $J^-(A)$, $D^-(S)$, $H^-(S)$ are defined
in a time dual manner.  $S$ is  a Cauchy surface  if and only if 
$D(S) : = D^+(S) \cup D^-(S) = M$, or equivalently,
$H(S):= H^+(S) \cup H^-(S) = \emptyset$.  If $M$ is 
globally hyperbolic, i.e., if $M$ has a Cauchy
surface $S$ then $M$ has topology $\Bbb R\times S$.

\subsection{Spacetimes of de Sitter type}
We use Penrose's notion of conformal infinity to to make precise
what it means for a spacetime to be asymptotically de Sitter.
We will say that $M$ has a {\it regular future conformal completion}
provided there  is a
spacetime-with-boundary $\tM$ with $C^2$ metric $\tg_{\alpha\beta}$ such that,
\begin{enumerate}
\item \label{point:interior}
$M$ is the interior of $\tM$, and hence $\tM = M \cup \scri$,
where $\scri = \partial \tM$,
\item \label{point:space}
$\scri$ is spacelike, and  $\scri\subset I^+(M,\tM)$, i.e.,
$\scri$ is the
future conformal boundary of $M$, and
\item $g_{\a\b}$ and $\tg_{\a\b}$ are related by,
$\tg_{\a\b}=\Omega^2g_{\a\b}$, where $\Omega\in C^2(\tM)$
satisfies: (i) $\Omega > 0$ on $M$ and (ii) $\Omega = 0$ and $d\Omega \ne
0$ along $\scri$.  (Then, since
$\scri$ is spacelike $\tD^{\gamma}\Omega$ must be timelike along $\scri$,
$\tD_{\gamma}\Omega \tD^{\gamma} \Omega
\onScri < 0$.)
\end{enumerate}

Similarly, we say that a spacetime $M$ has a \emph{regular past and future
conformal completion} if
the above definition holds, but with condition \ref{point:space}
modified as follows:
\begin{enumerate}
\item[\ref{point:space}$'$.] $\scri$ is spacelike, and decomposes into
disjoint nonempty sets,
$\scri = \scri^+\cup\scri^-$, where $\scri^+\subset I^+(M,\tM)$ and
$\scri^-\subset I^-(M,\tM)$, i.e.,
$\scri^+$ and $\scri^-$ are, respectively, the future and past conformal
boundaries of $M$.
\end{enumerate}

A spacetime admitting a regular future, or regular past and future,
conformal completion,
as described above, will be said to be of
\emph{de Sitter type}. In general, as a matter of notation, geometric
quantities associated with
$\tg_{\alpha\beta}$ will be decorated with a tilde~$\sim$, for example the
covariant derivative
$\tD_{\alpha}$ and Ricci tensor  $\tR_{\alpha\beta}$.

\subsection{Asymptotically simple spacetimes of de Sitter type}

Let $(M,g_{\alpha\beta})$ be a spacetime of de Sitter type with regular future
conformal infinity, $\scri^+$.
$M$ is said to be
\emph{future asymptotically simple} provided every future inextendible null
geodesic in $M$ has
 a future end point on $\scri^+$.  Future asymptotic
simplicity is a  global assumption which rules out the presence of
singularities,
black holes, etc.  Past asymptotic simplicity is defined time-dually.

The following proposition relates asymptotic simplicity
to the causal structure of $M$.

\begin{prop} \label{prop:structure}
Let $(M,g_{\alpha\beta})$ be a spacetime of de Sitter type with regular future
conformal infinity $\scri^+$.
\begin{enumerate}
\item \label{point:hypsimp}
If $M$ is globally hyperbolic and  $\scri^+$ is compact
then $M$ is future asymptotically simple.
\item \label{point:simphyp}
If $M$ is future asymptotically simple then $M$ is globally
hyperbolic.
\end{enumerate}
In either case, the Cauchy surfaces of $M$ are homeomorphic to
$\scri^+$.
\end{prop}
\begin{proof}

By extending $M\cup \scri^+$ a little beyond $\scri^+$, one
can obtain a spacetime without boundary $Q$ such that $\scri^+$
separates $Q$, and $Q= M\cup D^+(\scri^+, Q)$.
Suppose $M$ is globally hyperbolic and $\scri^+$ is compact.
Since any Cauchy surface for $M$ is clearly a Cauchy surface
for $Q$, $Q$ is globally hyperbolic.  Since  $\scri^+$ is compact,
it is necessarily a Cauchy surface for $Q$ (\cite{BILY,gal:cauchy}).  It
follows
that the Cauchy surfaces for $M$ are homeomorphic to $\scri^+$.
Let $\gamma$ be a null geodesic in $M$.
Since $\scri^+$ is a Cauchy surface for $Q$, the extension of $\gamma$
to the future in $Q$ must meet $\scri^+$.  It follows that
$M$ is future asymptotically simple. This proves point \ref{point:hypsimp}.

Now assume $M$ is future asymptotically simple.
We claim that $\scri^+$ is a Cauchy surface for $Q$.
Since, by construction, $H^+(\scri^+,Q) = \emptyset$,
we need only  show $H^-(\scri^+,Q)=\emptyset$. If
$H^-(\scri^+,Q) \ne \emptyset$, consider a future directed null
geodesic generator $\gamma$ of $H^-(\scri^+,Q)$.
By future asymptotic simplicity,
$\gamma$ meets $\scri^+$.  But a generator of $H^-(\scri^+,Q)$ can meet
$\scri^+$ only at an edge point of
$\scri^+$, yet $\scri^+$, being closed in $Q$, is edgeless. So we must have
$H^-(\scri^+,Q)=\emptyset$.

Thus, $\scri^+$ is a Cauchy surface for $Q$, and $Q$ is globally
hyperbolic.  It follows that $M$ can be foliated by Cauchy surfaces
for $Q$.  These Cauchy surfaces for $Q$ are  also
Cauchy surfaces for $M$,
and we conclude that $M$ is globally hyperbolic, with Cauchy surfaces
homeomorphic to $\scri^+$.  This proves point 2.
\end{proof}

\section{Past incomplete spacetimes of de Sitter type} \label{sec:past}
This section is devoted to the proof of the following two theorems, which
deal with spacetimes of de Sitter type with compact future conformal
infinity of nonpositive scalar curvature. The first theorem deals with
the case when $\scri^+$ has negative scalar curvature.
\begin{thm}\label{thm:incomplete}
Let $(M,g_{\alpha\beta})$ be a globally hyperbolic spacetime satisfying
the Einstein equations (\ref{eq:einst}) with
cosmological constant $\Lambda = n(n-1)/2$.
Assume
\begin{enumerate}
\item \label{point:desitter}
$M$ is of de Sitter type with future conformal
completion $(\tilde M, \tg_{\alpha\beta}, \tD_{\alpha} , \Omega)$,
with future conformal boundary $\scri^+$, which is  compact.
\item \label{point:energy}
The stress energy tensor $S_{\alpha\beta}$ of $(M,g_{\alpha\beta})$ satisfies
the strong energy condition (\ref{eq:strongenergy}) and the fall-off
condition,
\begin{equation}\label{eq:WEC-asympt}
\lim_{p \to \scri} \,
[ S_{\alpha\beta}\tD^{\alpha} \Omega  \tD^{\beta} \Omega](p) = 0 .
\end{equation}
\end{enumerate}
Then if $\scri^+$ has negative scalar curvature, every timelike geodesic in
$(M,g_{\alpha\beta})$ is past incomplete.
\end{thm}

Theorem \ref{thm:incomplete} may be viewed as a Lorentzian analogue of the
main result
of Witten-Yau \cite{witten:yau}.
The next theorem deals with the case when $\scri^+$ has zero scalar
curvature. This case is more subtle than the negative scalar curvature
case. Here we need to assume that both the strong and dominant energy
conditions hold, and in addition we require that $M$ is maximal.
\begin{thm} \label{thm:split}
Let $(M,g_{\alpha\beta})$ be a globally hyperbolic spacetime satisfying
the Einstein equations (\ref{eq:einst}) with
cosmological constant $\Lambda = n(n-1)/2$.
Assume that in addition to
conditions \ref{point:desitter} and \ref{point:energy} of Theorem
\ref{thm:incomplete}, the following holds.
\begin{enumerate}
\setcounter{enumi}{2}
\item \label{point:dec}
$(M,g_{\alpha\beta})$ satisfies the dominant energy condition.
\item $(M,g_{\alpha\beta})$ is maximal among all spacetimes satisfying
conditions \ref{point:desitter}--\ref{point:dec}.
\end{enumerate}
Then, if $\scri^+$ has zero scalar curvature,
either $(M,g_{\alpha\beta})$ contains no
past complete timelike geodesic,
or else $(M,g_{\alpha\beta})$ is isometric
to the warped product with line element
\begin{equation}\label{eq:split}
ds^2 = - d\tau^2 + e^{2\tau} h_{ab} dx^a dx^b
\end{equation}
with $h_{ab}$ a Ricci flat metric on $\scri^+$. In particular, $(M,
g_{\alpha\beta})$ satisfies the vacuum Einstein equations with cosmological
constant $\Lambda = n(n-1)/2$.
\end{thm}
\begin{remark}
The condition (\ref{eq:WEC-asympt}) may be weakened to
\begin{align*}
\lim_{p \to \scri} \,
[ \Omega^2 S_{\alpha\beta}\tD^{\alpha} \Omega  \tD^{\beta} \Omega](p) &= 0 \\
\liminf_{p \to \scri} \,
[  S_{\alpha\beta}\tD^{\alpha} \Omega  \tD^{\beta} \Omega](p) &\geq 0
\end{align*}
\end{remark}

\begin{remark}\label{rem:scri}
It follows from Proposition \ref{prop:structure} that the Cauchy surfaces of
$M$ are homeomorphic to $\scri^+$, and in particular are compact.
\end{remark}

\begin{remark}
Let $N^n$ be a smooth compact manifold of dimension $n\ge 3$.  By
definition, $N$ is of Yamabe type $-1$, if $N$
admits a metric of constant negative scalar curvature, but not of  zero or
constant positive curvature; $N$ is of
Yamabe type $0$ if $N$ admits a metric of zero scalar curvature, but not
constant positive scalar curvature; $N$ is
of Yamabe type
$+1$ if $N$ admits a metric of constant positive curvature.  The definition
of Yamabe type partitions the class of
all compact $n$-manifolds into these three sub-classes.  Thus, according to
Theorem~\ref{thm:incomplete},
if $\scri^+$ is of Yamabe type $-1$ then all timelike geodesics in $M$ are
past incomplete.
Or, to change the
viewpoint slightly, if $M$ contains a past complete timelike geodesic,
then the
Yamabe type of $\scri^+$
must be $0$ or $+1$, and  is
$0$, only if $M$ splits as a warped product, as described.
\qed
\end{remark}

Theorems \ref{thm:incomplete} and \ref{thm:split}
shall be obtained as consequences of the following basic
singularity theorem, and a rigid version of it,
for spacetimes obeying the energy condition (\ref{eq:strongenergy}).

\begin{prop}\label{prop:sing}
Let $M^{n+1}$ be a spacetime satisfying the energy condition,
$$
{\rm Ric}\,(V,V) = R_{\alpha\beta}V^{\alpha}V^{\beta} \ge -n
$$
for all unit timelike vectors $V^{\alpha}$.  Suppose that $M$ has a smooth
compact Cauchy surface $N$ with
mean curvature $H$ satisfying $H> n$.  Then every timelike geodesic in $M$
is past incomplete.
\end{prop}
\begin{remark}
By our sign conventions, $H = {\rm div}_N\, T= D_{a}T^a$, where $T$ is the
future pointing unit normal
along $N$.  Proposition \ref{prop:sing} is an extension of an old
singularity theorem of Hawking
to the case of negative Ricci curvature.
\end{remark}
\begin{proof}
Fix $\delta > 0$ so that the mean curvature of $N$ satisfies $H\ge
n(1+\delta)$.  Let
$\rho : I^-(N) \to \Re$ be the Lorentzian distance function to $N$,
\beq
\rho(x) = d(x,N) = \sup_{y\in N}d(x,y)  \,;
\eeq
$\rho$ is continuous and smooth outside the past focal cut locus of $N$.
We will show that $\rho$ is bounded from
above,
\beq
\rho(x) \le \coth^{-1}(1+\delta) \qquad \mbox{for all } x\in I^-(N)  \,.
\eeq
This implies that every past inextendible timelike curve with future end
point on $N$ has
length $\le \coth^{-1}(1+\delta)$.

Suppose to the contrary, there is a point $q\in I^-(N)$ such that
$d(q,N)=\ell> \coth^{-1}(1+\delta)$.
Let $\gamma:[0,\ell]\to M$, $t\to\gamma(t)$, be a past directed unit speed
timelike geodesic from
$p\in N$ to $q$ that realizes the distance from $q$ to $N$. $\gamma$ meets
$N$ orthogonally, and,
because it maximizes distance to $N$, $\rho$ is smooth on an open set $U$
containing
$\gamma\setminus \{q\}$.  For $0\le t< \ell$, the slice $\rho = t$ is
smooth near
$\gamma(t)$; let $H(t)$ be the mean curvature, with respect to the
future pointing normal $D^{\alpha}\rho$, at $\gamma(t)$ of the slice $\rho
= t$.

$H=H(t)$ obeys the traced Riccati (Raychaudhuri's) equation,
\beq\label{eq:ray}
H' = {\rm Ric}(\gamma',\gamma') +|K|^2\,,
\eeq
where $'=d/dt$ and $|K|^2=K_{ab}K^{ab}$ is the square of the second
fundametal form
$K_{ab}$ of $N$.
Equation (\ref{eq:ray}), together with the inequalities  $|K|^2 \ge (\tr
K)^2/n = H^2/n$,
${\rm Ric}(\gamma',\gamma')\ge -n$ and $H(0) \ge n(1+\delta)$, implies that
$\mathscr H(t):=H(t)/n$
satisfies,
\beq
\mathscr H' \ge \mathscr H^2 - 1, \qquad \mathscr H(0) \ge 1+\delta\,.
\eeq
By an elementary comparison with the unique solution to: $h' = h^2 -1$,
$h(0) = 1+\delta$, we
obtain $\mathscr H(t)\ge \coth(a-t)$, where $a = \coth^{-1}(1+\delta) <
\ell$, which implies
that $\mathscr H = \mathscr H(t)$ is unbounded on $[0,a)$, contradicting
the fact that $\mathscr H$ is
smooth on $[0,\ell)$.
\end{proof}

Proposition \ref{prop:sing} admits the following rigid generalization.
\begin{prop}\label{prop:rigsing}
Let $M^{n+1}$ be a spacetime satisfying the energy condition,
\beq
{\rm Ric}\,(V,V) = R_{\alpha\beta}V^{\alpha}V^{\beta} \ge -n
\eeq
for all unit timelike vectors $V^{\a}$, and suppose $M$ has a smooth
compact Cauchy surface $N$ with
mean curvature $H$ satisfying $H\ge n$.  If there exists at least one  past
complete timelike geodesic in $M$,
then a neighborhood of $N$ in $J^-(N)$ is isometric to $(-\epsilon,0]\times
N$, with warped product
metric $ds^2 = - d\tau^2 + e^{2\tau} h_{ab} dx^a dx^b$, where $h_{ab}$ is
the induced metric on $N$.
If the timelike geodesics orthogonal to $N$ are all past complete, then
this warped product splitting
extends to all of $J^-(N)$.
\end{prop}
\begin{proof}
The proof method we employ is standard.  Let $h_{ab}$ be the induced metric
on $N$, and let
$K_{ab}$ be the second fundamental form of $N$, $K_{ab} = -D_aT_b$,  where
$T$ is the past pointing unit
normal along~$N$.

Let $t\to N_t$ be a variation of $N_0 = N$, with variation
vector field $\phi T$, where $\phi$ is a smooth function on $N$.
Let $H = H_t$ be the mean curvature function of $N_t$.
A standard computation gives
\beq\label{eq:var}
\frac{\partial H}{\partial t} \big |_{t=0}
= - \triangle \phi + (R_{TT} + \frac{H_0^2}{n}+ \sigma_{ab}\sigma^{ab})\phi \,,
\eeq
where $R_{TT} = {\rm Ric}(T,T)$, and  $\sigma_{ab}$ is the trace free part
of $K_{ab}$,
$\sigma_{ab} = K_{ab}-\frac{H_0}{n}h_{ab}$.
In view of the energy condition and the fact that $H_0 \ge n$, the quantity
$R_{TT} + \frac{H_0^2}{n}+
\sigma_{ab}\sigma^{ab}$ is nonnegative. If it were positive at some point
then, by standard results, there would
exist a function $\phi$ for which the right hand side of
(\ref{eq:var}) were strictly positive.  Since $H_0\ge n$, this would imply
that for small $t>0$, $H_t > n$.
Proposition \ref{prop:sing}  would then imply that all timelike geodesics
are past incomplete, contrary to
assumption.   Thus, $\sigma_{ab}$ must vanish along $N$, and hence $N$ is
totally umbilic, $K_{ab} = h_{ab}$ and
$H=n$.

Now introduce Gaussian normal coordinates in a neighborhood  $U$ of $N$ in
$J^-(N)$,
\beq
U = [0,\epsilon)\times N, \quad ds^2 = -du^2 +h_{ab}(u)dx^adx^b \,.
\eeq
Let $K_{ab} = K_{ab}(u)$ and $H=H_u$ be the second fundamental form and
mean curvature,
respectively, of the $u$-slice $N_u$.
$H=H_u$ obeys the traced Riccati equation,
\beq
\frac{\partial H}{\partial u} = R_{uu} + |K|^2  \,,
\eeq
where $R_{uu}$ is the Ricci tensor contracted with the coordinate vector
$\partial_u$.
Since $R_{uu} \ge -n$, $|K|^2 \ge H^2/n$, and $H_0 = n$, it follows  that
$\mathscr H :=H/n$ satisfies,
\beq
\frac{\partial \mathscr H}{\partial u} \ge \mathscr H^2 -1,\quad \mathscr
H(0) = 1 \,,
\eeq
which by an elementary comparison, implies $\mathscr H \ge 1$ on $U$.
Hence, $H|_{N_u} \ge n$ for all $u\in [0,\epsilon)$.
But the argument above then implies that each $N_u$ is totally umbilic,
$K_{ab}(u)  = h_{ab}(u)$
for each $u$.  Since $K_{ab} = -\frac12\frac{\partial h_{ab}}{\partial u}$,
we obtain
the warped product splitting of $U$,
\beq
ds^2  = -du^2 + e^{-2u}h_{ab}(0)dx^adx^b  \,,
\eeq
which, upon the substitution $\tau =-u$, yields the local warped product
splitting asserted in the
proposition.
If the normal geodesics to $N$ are all past complete then this splitting
can be extended indefinitely to
the past.
\end{proof}

We now proceed to the proofs of Theorems \ref{thm:incomplete} and
\ref{thm:split}.

\begin{proof}[Proof of Theorem \ref{thm:incomplete}]
We will begin by proving that (\ref{eq:WEC-asympt}) implies $\tD_{\gamma}
\Omega \tD^{\gamma} \Omega \onScri$ $= -1$. To see this, note that
$\tD^{\alpha}\Omega \onScri$ is perpendicular to $\scri^+$, and is past
oriented.
Introduce a coordinate system $(x^{\alpha}) = (s,x^a)$
near $\scri^+$ so that $\partial_s \onScri$ agrees with $\tD^{\alpha} \Omega
\partial_{x^{\alpha}} \onScri$. Then
$Y^{\alpha}  = \Omega (\partial_s)^{\alpha} = \Omega \tD^{\alpha} \Omega +
O(\Omega^2)$.
The Ricci tensor $R_{\alpha\beta}$ of $(M,g_{\alpha\beta})$ and the Ricci
tensor $\tR_{\alpha\beta}$ of $(M, \tg_{\alpha\beta})$ are related by
\begin{equation}\label{eq:ConfRic}
R_{\alpha\beta} =
\tR_{\alpha\beta} + \Omega^{-1} [ (n-1) \tD_{\alpha}\tD_{\beta}\Omega
+ \tD_{\gamma} \tD^{\gamma} \Omega \tg_{\alpha\beta}]
- n \Omega^{-2} \tD_{\gamma} \Omega \tD^{\gamma} \Omega \tg_{\alpha\beta}
\end{equation}
A computation shows that
$$
S_{\alpha\beta} Y^\alpha Y^\beta =
\left [ \frac{n(n-1)}{2} \tD_{\gamma} \Omega \tD^{\gamma} \Omega + \Lambda
\right ] g_{\alpha\beta} Y^{\alpha} Y^{\beta} + O(\Omega)
$$
and hence (\ref{eq:WEC-asympt}) implies
\begin{equation}\label{eq:DOmScri}
\tD_{\gamma} \tD^{\gamma} \Omega\onScri =
-1 .
\end{equation}

Let $\th^0_{ab}$ be the metric on $\scri^+$ induced from
$\tg_{\alpha\beta}$.
By the Yamabe theorem, there is a positive
function $\theta$ on
$\scri^+$ such that the scalar curvature
$\br^0$ of $\theta^{-2}\th^0_{ab}$ equals $-1$, $0$, or $1$ on $\scri^+$.
Further, there is a neighborhood $U$ of $\scri^+$, and a
conformal gauge transformation $\Theta$ with $\Theta
\onScri = \theta$, such that after replacing $\Omega$ by $\bar \Omega = \Omega
\Theta^{-1}$, and $\tilde g_{\alpha\beta}$ by $\bar g = \Theta^{-2}\tilde
g_{\alpha\beta}$, we have
$$
\Omega(p) = d_{\bar g} (p, \scri^+)
$$
on $U$ where $d_{\bar g}$ denotes
the Lorentz distance to $\scri^+$. This is achieved, following
\cite[\S 5]{andersson:dahl} by solving the equation
\begin{equation}\label{eq:gausscond}
- 1 = \bg^{\alpha\beta} D_{\alpha} \bOmega D_{\beta} \bOmega
\end{equation}
By (\ref{eq:DOmScri}) the function $a = \Omega^{-1} ( 1+
\tg^{\alpha\beta}D_{\alpha} \Omega D_{\beta} \Omega )$ is in $C^1(\tM)$. A
computation shows that (\ref{eq:gausscond}) is equivalent to the system
$$
2 \Theta \tg^{\alpha\beta} D_{\alpha} \Theta D_{\beta} \Omega -
\Omega \tg^{\alpha\beta} D_{\alpha} \Theta D_{\beta} \Theta = \Theta^2 a
$$
This equation with initial data $\Theta = \theta$ on $\scri^+$, has a unique
solution in a neighborhood of $\scri^+$ \cite[pp. 39-40]{spivak:V}.
In a sufficiently small neighborhood $U$ of $\scri^+$, the solution is
positive, and we continue this to a positive function $\Theta$ on all of $M$
for which $\bg^{\alpha\beta} D_{\alpha}\bOmega D_{\beta} \bOmega = -1$ on
$U$. This implies that the gradient curves of $\bOmega$ on $U$ are unit-speed
timelike geodesics with respect to $\bg$ and since $\bOmega = 0$ on $\scri^+$,
we have $\bOmega(p) = d_{\bg}(p,\scri^+)$. Finally, we rename
$\bg_{\alpha\beta}, \bOmega$ to $\tg_{\alpha\beta}, \Omega$.
By construction, the metric $\th^0_{ab}$ induced on
$\scri^+$ by $\tg_{\alpha\beta}$ has scalar curvature $\tilde r^0 = \br^0$.

Letting $t = \Omega$, so that $t$ {\em
increases to the past} near $\scri^+$, the foliation
of level sets $N_t$ of $t$ is the Gaussian foliation with respect to
$\scri^+$ on $U$. Let $h_{ab}, r, K_{ab} , H$ be the induced metric on $N_t$,
its scalar curvature function, the second fundamental form of $N_t$ and the
mean curvature $H = h^{ab} K_{ab}$,  respectively. Here
$K_{ab} = - D_a T_b$ is the second fundamental form of $N_t$ defined with
respect to the past directed timelike normal $T$ to $N_t$, so that $K_{ab} =
- \half \Lie_T g_{ab}$. Similarly,
$\th_{ab}, \tilde r, \tK_{ab}, \tH$ are the metric, scalar curvature, second
fundamental form, and mean curvature of $N_t$,
defined with respect to the conformally rescaled metric
$\tg_{\alpha\beta}$.

By the above, we may without loss of generality assume
that on $U$,
$g_{\alpha\beta}$ is of the form
\begin{equation}\label{eq:gform}
g_{\alpha\beta} dx^{\alpha} dx^{\beta}  = \frac{1}{t^2} ( - dt^2 + \th_{ab}
dx^a dx^b )
\end{equation}
where $\th_{ab} = \th_{ab}(t,x)$, $\th_{ab}(0,x) = \th^0_{ab}(x)$ and the
scalar curvature $\tilde r^0$ of $\th^0_{ab}$ is constant $= -1,0,+1$.
With this form for $g_{\alpha\beta}$ and $\tg_{\alpha\beta}$, we have
$T = t \partial_t$. We will use an index $T$ to denote contraction with $T$,
for example $u_T = u_{\alpha} T^{\alpha}$, and an index $0$ for contraction
with $\partial_t$, for example $u_0 = u_{\alpha} (\partial_t)^{\alpha}$.

To prove Theorem \ref{thm:incomplete}, it is sufficient to show,
assuming $\tilde r_0 = -1$,
that $H|_{N_t}>n$ for some $t>0$, for then Proposition \ref{prop:sing} applies.
The mean curvature functions $H$ and $\tH$ are related by
\beq\label{eq:meancurv}
H = t \tH + n \,.
\eeq
In particular, $H \onNt > 0$ for $t$ sufficiently small.  The Gauss
equation (in the physical
metric $g_{\alpha\beta}$) applied to each
$N_t$, together with the Einstein equation, yields the constraint,
\beq
H^2 &=& 2S_{TT} +2\Lambda +|K|^2 - r \nonumber \\
&=& 2t^2S_{00} + n(n-1) +|K|^2 - t^2 \tilde r \,,
\eeq
which, since $|K|^2 \ge H^2/n$,  implies,
\beq
H^2 \ge n^2 + \frac{n}{n-1}t^2(S_{00} -\tilde r)  \,.
\eeq
Using $\tilde r_0=-1$ and the energy condition (\ref{eq:WEC-asympt}), the
above inequality
implies $H|_{N_t} > n$ for all $t>0$ sufficiently small.
\end{proof}

Finally, we give the proof of Theorem \ref{thm:split}.
\begin{proof}[Proof of Theorem \ref{thm:split}]
Let the notation be as in the proof of Theorem \ref{thm:incomplete}.
We will show, by adapting an argument in \cite{Anderson}
to the Lorentzian
setting, that $H|_{N_t} \ge n$ for each $t>0$,  so that Propsition
\ref{prop:rigsing}
may be applied.
To this end, we first show that the quantity $t^{-1}\tH$ is nondecreasing
along the
flow lines of $\partial_t$.

The conformal transformation rule for the Ricci tensor, shows that on $U$,
\begin{subequations}\label{eq:confcurv}
\begin{align}
R_{00} &= \tR_{00} - t^{-1} \tH - t^{-2}n  \label{eq:confR00} \\
R &=  n(n+1) + 2nt \tH + t^2 \tR \label{eq:ConfR}
\end{align}
\end{subequations}
The traced Riccati (Raychaudhuri) equation in the conformally rescaled
metric $\tg_{\alpha\beta}$
is
\begin{equation}\label{eq:dttildeH}
\partial_t \tH = |\tK|^2 + \tR_{00} .
\end{equation}
Using (\ref{eq:confR00}), (\ref{eq:dttildeH})
gives,
\beq\label{eq:dtH}
\partial_t (t^{-1} \tH)  & = & t^{-1} [|\tK|^2 + t^{-2} (R_{TT} + n)] \\
&\ge & 0 \, ,
\eeq
since, by the energy condition (\ref{eq:strong-alt}), $R_{TT} + n\ge 0$.
Integrating  from $\epsilon$ to $t$ along each flow line
of $\partial_t$, and letting $\epsilon \to 0$, gives,
\beq\label{eq:liminf}
t^{-1}\tH(t)} \ge \liminf_{\epsilon\to 0} {\epsilon^{-1}\tH(\epsilon) \,.
\eeq

The contracted Gauss equation (in the unphysical metric
$\tg_{\alpha\beta}$) states
$$
\tilde r + \tH^2 - |\tK|^2 = 2 \tR_{00} + \tR  \,.
$$
Using (\ref{eq:confcurv}) we find after some manipulations, that on $U$,
$$
2 \tR_{00} + \tR = 2 t^{-2} S_{TT} - 2(n-1) t^{-1} \tH
$$
The previous two equations imply,
$$
2(n-1) t^{-1}\tH \left (1 + \frac{t \tH}{2(n-1)}\right )
= |\tK|^2 - \tilde r + 2 t^{-2} S_{TT} \,.
$$
Setting $t = \epsilon$ in the above, and letting $\epsilon\to 0$, while
making use of the  asymptotic form of the weak energy condition
(\ref{eq:WEC-asympt}), and the fact that $\tH$ is bounded, shows that
the right hand side of Equation~(\ref{eq:liminf}) is greater than or
equal to $-\tilde r_0/2(n-1)$.  Since we are now considering the case
$\tilde r_0 = 0$,
we conclude that there exists $c > 0$ sufficiently small so that
$\tH|_{N_t} \ge 0$ for $t \in (0,c]$.  Equation
(\ref{eq:meancurv}) then implies
$H|_{N_t} \ge n$ for each $t \in (0,c]$. Let $M_c = \{ p \in M, t(p) \in
(0,c]\}$.

By applying Proposition \ref{prop:rigsing} to $N_t \subset M$ for $t\in
(0,c]$,
we conclude that the line element on $M_c$
is of the form (\ref{eq:split}) and hence is conformal to
$-dt^2 +  \th_{ab} dx^a dx^b$. Taking into account the assumption that
$\scri^+$ is a regular conformal boundary, it follows that $\th_{ab} =
\th^0_{ab}$, the metric on $\scri^+$.
Hence, for $t \in (0,c]$,
$N_t$ with induced metric
$h_{ab}$ has zero scalar
curvature and the second fundamental form of $N_t$ satisfies $K_{ab} =
h_{ab}$.
The Hamiltonian constraint
$$
r + (\tr K)^2 - |K|^2 = 2 \Lambda + S_{TT}
$$
then implies
$$
S_{TT} = 0
$$
By assumption the dominant energy condition holds, and hence by the
conservation theorem \cite[\S 4.3]{HE}, $S_{\alpha\beta} = 0$ in the domain
of dependence of $N_c$. By construction, $N_c$ is a Cauchy surface in $M$, so
we find that $S_{\alpha\beta} = 0$ in $M$. A calculation shows that a warped
product line element of the form (\ref{eq:split}) satisfies the vacuum
Einstein equations with cosmological constant $\Lambda = n(n-1)/2$ only if
$\Ric[\th_{ab}^0] = 0$.

Summarizing our conclusions so far,
we have that $(M,g_{\alpha\beta})$ satisfies the vacuum Einstein equations
with cosmological constant $\Lambda$ and we have a Cauchy surface $N_c$ in
$M$ with induced data equivalent to that of a hypersurface in the warped
product spacetime with line element given by (\ref{eq:split}).
By assumption $(M,g_{\alpha\beta})$ is maximal.
The global splitting asserted in Theorem
\ref{thm:split} is a consequence of
uniqueness of the maximal Cauchy development for the Einstein equations,
see \cite{ChB:geroch} for a discussion of the $\Lambda = 0$ case.
\end{proof}

\section{Spacetimes of de Sitter type and the null energy condition}
\label{sec:simp}

In this section we obtain restrictions on the topology of spacetimes
of de~Sitter type which obey the null energy condition.

\begin{thm}\label{thm:null}
Let $(M,g_{\a\b})$ be a spacetime of
de Sitter type with regular past and future conformal boundaries $\scri^{\pm}$.
Assume that $(M,g_{\a\b})$ is future or past asymptotically simple and
satisfies the null
energy condition (\ref{eq:NEC}). Then $(M,g_{\a\b})$ is globally
hyperbolic, and the Cauchy surfaces of $M$
are compact with finite fundamental group.
\end{thm}
\begin{remark}
Put another way, given a globally hyperbolic spacetime $M$ of de Sitter
type with regular $\scri^{\pm}$,
which obeys the null energy condition, if  the fundamental group of the
Cauchy surfaces of $M$ is infinite
then $M$ can be neither past nor future asymptotically simple.  This is
illustrated by the spatially closed
version of Schwarzschild-de Sitter spacetime,
whose Cauchy surfaces have
topology $S^1\times S^2$.  Asymptotic simplicity fails in this model due
to the presence of a black hole and a white hole.
\end{remark}
\begin{proof} For definiteness, assume $M$ is future asymptotically simple.
By Proposition \ref{prop:structure}, $M$ is globally hyperbolic, with Cauchy
surfaces homeomorphic to $\scri^+$.
Then $\tilde M$ can be extended a little
beyond $\scri^{\pm}$ to obtain a spacetime (without boundary) $P$, with
$\tilde M\subset P$,
such that the Cauchy surfaces for $M$ are also Cauchy surfaces for $P$, so
that $P$
is globally hyperbolic.

Consider the achronal boundary $\partial I^+(p,P)$, which is an achronal
$C^0$ hypersurface in
$P$. We claim that  $\partial I^+(p,P)$ is compact.  Suppose not.  By the
global hyperbolicity
of $P$, $\partial I^+(p,P) = J^+(p,P) \setminus I^+(p,P)$, and hence the
null geodesic generators
of $\partial I^+(p,P)$ extend back to the point $p$. Using the
noncompactness of $\partial I^+(p,P)$, one easily constructs
a future directed null geodesic $\gamma\subset \partial I^+(p,P)$ starting
at $p$, which is future
inextendible in $P$.  In particular, $\gamma$ meets $\scri^+$ at a point
$q$, say, and enters
the interior of $D^+(\scri^+,P)$.

Let $\eta$ be the portion of $\gamma$ from $p$ to $q$, excluding these end
points.  Then  $\eta$
is a \emph{null line}, i.e., a complete achronal null geodesic in
$(M,g_{\a\b})$.
Observe that $I^+(\eta, M) = I^+(p,P) \cap M$, from which it follows that
$\partial I^+(\eta,M)
= \partial I^+(p,P)\cap M$ (where $\partial I^+(A,X)$ refers to the
boundary in $X$).  It follows that
the generators of the achronal boundary $\partial I^+(\eta,M)$ extend back
to $p$ and hence are
past complete in $(M,g_{\a\b})$.  By the time-dual of these arguments, we
have that $\partial I^-(\eta,M)
=\partial I^-(q,P)\cap M$, and that the null generators of $\partial
I^-(\eta,M)$ are future complete
in $(M,g_{\a\b})$.
Then, since the null energy condition holds, we
may apply the  null splitting theorem \cite{galloway:nullsplit} to conclude
that $\partial
I^+(\eta,M)$ and $\partial
I^-(\eta,M)$ agree, and, in fact, form a smooth achronal edgeless totally
geodesic null hypersurface in $M$.
Hence, $\partial I^+(p,P)\cap M = \partial I^-(q,P)\cap M$, from which it
follows that the null generators of
$\partial I^+(p,P)$ reconverge, and, by the achronality of $\partial
I^+(p,P)$, terminate at $q$.  But this
contradicts the fact that the generator $\gamma$ enters the interior of
$D^+(\scri^+,P)$.

Thus, $\partial I^+(p,P)$ is compact, and, by standard results
\cite{BILY,gal:cauchy}, is a Cauchy
surface for $P$.  So the Cauchy
surfaces of $P$, and hence, the Cauchy surfaces of $M$ are compact.  Now
let $M^*$ denote the universal
covering spacetime of $M$.  $M^*$ will be globally hyperbolic; in fact if $S$
is a Cauchy surface for $M$, so that $M\approx \Re \times S$, then
$M^*\approx \Re \times S^*$, where $S^*$ is the universal cover of $S$
and each slice $\{t\}\times S^*$
is a Cauchy surface for $M^*$.  It is easily seen that the assumptions on
$M$ in the theorem lift to $M^*$.
Then from the above, we conclude that the Cauchy surfaces of $M^*$, and
hence $S^*$, are compact.  It follows
that  $S^*$ is a finite cover of $S$, and hence $S$ has a finite
fundamental group.
\end{proof}

The following corollary to Theorem \ref{thm:null}, is an immediate
consequence of Theorem \ref{thm:null}, and
Proposition \ref{prop:structure}.
It replaces
the assumption
of asymptotic simplicity with other natural assumptions.

\begin{cor}\label{cor:null}
Let $(M,g_{\alpha\beta})$ be a globally hyperbolic spacetime of
de Sitter type with regular past and future conformal boundaries $\scri^{\pm}$.
Assume that $(M,g_{\alpha\beta})$ obeys the null energy condition, and that
$\scri^+$ (or  $\scri^-$) is compact.
Then  the Cauchy surfaces
of $M$, which by Proposition \ref{prop:structure} are homeomorphic to
$\scri^+$ (or $\scri^-$), have finite fundamental group.
\end{cor}

We conclude with the following theorem.

\begin{thm} \label{thm:worm}
Let $(M^{n+1},g_{\a\b})$, $n\le 7$, be a globally hyperbolic
spacetime of de Sitter type with regular
future conformal boundary $\scri^+$. Assume that
$(M^{n+1},g_{\a\b})$ obeys the null energy condition, and that $\scri^+$ is
compact and orientable.
If $M$ is past null geodesically complete then the Cauchy surfaces of $M$,
which by Proposition
\ref{prop:structure}, are homeomorphic to $\scri^+$,
 have vanishing co-dimension
one homology,
i.e., $H_{n-1}(N,\mathbb Z) = 0$, $N$ a Cauchy surface for $M$. In particular,
there can be no worm holes in~$N$.
\end{thm}

\begin{proof} The proof is  an application of the Penrose singularity
theorem \cite{HE,ON} applied
to a suitable covering spacetime of $M$.

As in the proof of Theorem \ref{thm:incomplete}, introduce coordinates so
that the physical
metric $g_{\a\b}$ takes the form of Equation (\ref{eq:gform}).  The second
fundamental forms
$K_{ab}$  and $\tK_{ab}$, of the $t$-slices $N_t$ (notation as in the proof of
Theorem \ref{thm:incomplete}) are related by
\beq\label{eq:fundform}
K_{ab} =  t^{-1} \tK_{ab} + g_{ab}  \,.
\eeq
Let $\tilde X$ be a $\tg$-unit vector field defined in a neighborhood $U$
of a point $p\in\scri^+$,
which is everywhere orthogonal to $\partial_t$.  Then $X=t^2 \tilde X$ is a
$g$-unit vector field
defined on $U\setminus \scri^+$, everywhere orthogonal to $\partial_t$.
From (\ref{eq:fundform}), we have
\beq
K_{ab}X^aX^b = t\,\tK_{ab}\tilde X^a\tilde X^b + 1\,,
\eeq
which is positive for $t$ sufficiently small. Hence, for $t$ sufficiently
small,
$K_{ab} = -D_aT_b$ is positive definite along $N_t$.

Thus, by fixing $t_0$ sufficiently small, there exists a compact Cauchy
surface $N=N_{t_0}$ for $M$
which is strictly convex to the past, i.e. for which $D_aT_b$ is negative
definite along $N$, where $T$ is the
past pointing unit normal along $N$.  Suppose $H_{n-1}(N,\mathbb Z) \ne 0$.
By well known
results of geometric measure theory (see \cite[p. 51]{lawson}, for
discussion),
every nontrivial class in $H_{n-1}(N,\mathbb Z)$ has a least area
representative
which can be expressed as  a linear combination of smooth, orientable,
connected, compact,
embedded minimal (mean curvature zero) hypersurfaces in $N$.  Let $\Sigma$
be such a
hypersurface; we may assume  $\Sigma$ represents a nontrivial element of
$H_{n-1}(N,\mathbb Z)$.
Note $\Sigma$ is spacelike
and has co-dimension two in
$M$.   As described in \cite{gal:min}, since $\Sigma$ is minimal in $N$,
and $N$
is strictly
convex to the past in $M$, $\Sigma$ must be a past trapped
surface in
$M$, i.e., the two families of past directed  null geodesics issuing
orthogonally from $\Sigma$ are converging in
the mean along $\Sigma$.

The next step is to construct a certain covering spacetime $M^*$.  Since
$N$ is a Cauchy surface for $M$,
so that $M\approx \Re \times N$, each covering space $N^*$ of $N$ gives
rise, in an essentially unique
way, to a covering spacetime $M^*$ of $M$, such that $M^*\approx \Re \times
N^*$, where the slices
$\{t\}\times N^*$ are Cauchy surfaces for $M^*$.  Since $\Sigma$
is
two-sided, loops in $N$ have a
well-defined oriented intersection number
with respect to $\Sigma$.  The
intersection number
is a homotopy invariant,
and so gives rise to a well-defined subgroup $G$
of $\Pi_1(N)$,
corresponding
to the loops in $N$ having zero intersection number with
respect to
$\Sigma$.  $N^*$ is defined to be
the covering space of $N$
associated with the subgroup $G$, i.e., satisfying
$\pi_*(\Pi_1(N^*)) = G$,
where $\pi:N^* \to N$ is the covering map.  $N^*$
has a simple
description
in terms of cut-and-paste operations.  $\Sigma$ does not
separate $N$, for
otherwise it would bound
in $N$.  By making a cut along
$\Sigma$, we obtain a compact manifold $N'$
with two boundary
components,
each isometric to $\Sigma$.  Taking $\mathbb Z$ copies of $N'$,
and gluing
these copies end-to-end
we obtain the covering space $N^*$ of
$N$.  The inverse image
$\pi^{-1}(\Sigma)$ consists
of $\mathbb Z$ copies of
$\Sigma$, each one separating $N^*$.  Let
$\Sigma_0\subset N^*$ denote one
of these
copies.

As per the discussion above, there exists a covering spacetime $M^* \approx
\Re\times N^*$,
with Cauchy surfaces homeomorphic to $N^*$. Since the covering map is a
local isometry,
the assumptions that $M$ obeys the null energy condition and is past null
geodesically
complete lift to $M^*$.  Moreover, $\Sigma_0$ will be a past trapped
surface in $M^*$.
Then, according to the Penrose singularity theorem (cf., \cite[Theorem
1]{HE} or \cite[Theorem 61]{ON}),
the achronal boundary $\partial I^-(\Sigma_0)$ is a \emph{compact} Cauchy
surface for $M^*$.
This implies that $N^*$ is compact, and hence, a finite covering of $N$,
which is a contradiction.
We conclude that $H_{n-1}(N,\mathbb Z) = 0$.
\end{proof}

\noindent{\bf Acknowledgements:} We thank Mike Anderson and Helmut Friedrich
for useful comments and discussion. Theorems 3.1 and 3.2 were proved
independently, with somewhat different proofs, by Mike Anderson
\cite{anderson:priv}.


\begin{thebibliography}{10}

\bibitem{Anderson}
M.~T. Anderson, \emph{Boundary regularity, uniqueness and non-uniqueness for
  {AH} einstein metrics on 4-manifolds}, math.DG/0104171, 2001.

\bibitem{anderson:prescribed}
\bysame, \emph{Einstein metrics with prescribed conformal infinity on
  4-manifolds}, math.DG/0105243, 2001.

\bibitem{anderson:priv}
\bysame, \emph{private communication}, 2002.

\bibitem{andersson:dahl}
Lars Andersson and Mattias Dahl, \emph{Scalar curvature rigidity for
  asymptotically locally hyperbolic manifolds}, Ann. Global Anal. Geom.
  \textbf{16} (1998), no.~1, 1--27.

\bibitem{perlmutter}
N.~Bahcall, J.P. Ostriker, S.~Perlmutter, and P.~J. Steinhardt, \emph{The
  cosmic triangle: revealing the state of the universe}, Science \textbf{284}
  (1999), 1481--1488.

\bibitem{BILY}
Robert Budic, James Isenberg, Lee Lindblom, and Philip~B. Yasskin, \emph{On
  determination of {C}auchy surfaces from intrinsic properties}, Comm. Math.
  Phys. \textbf{61} (1978), no.~1, 87--95.

\bibitem{cai:galloway}
Mingliang Cai and Gregory~J. Galloway, \emph{Boundaries of zero scalar
  curvature in the {A}d{S}/{C}{F}{T} correspondence}, Adv. Theor. Math. Phys.
  \textbf{3} (1999), no.~6, 1769--1783 (2000).

\bibitem{ChB:geroch}
Yvonne Choquet-Bruhat and Robert Geroch, \emph{Global aspects of the {C}auchy
  problem in general relativity}, Comm. Math. Phys. \textbf{14} (1969),
  329--335.

\bibitem{friedrich:positive}
Helmut Friedrich, \emph{Existence and structure of past asymptotically simple
  solutions of {E}instein's field equations with positive cosmological
  constant}, J. Geom. Phys. \textbf{3} (1986), no.~1, 101--117.

\bibitem{friedrich:EYM}
\bysame, \emph{On the global existence and the asymptotic behavior of solutions
  to the {E}instein-{M}axwell-{Y}ang-{M}ills equations}, J. Differential Geom.
  \textbf{34} (1991), no.~2, 275--345.

\bibitem{GSWW}
G.~J. Galloway, K.~Schleich, D.~M. Witt, and E.~Woolgar, \emph{Topological
  censorship and higher genus black holes}, Phys. Rev. D (3) \textbf{60}
  (1999), no.~10, 104039, 11.

\bibitem{GSWW2}
\bysame, \emph{The {A}d{S}/{C}{F}{T} correspondence and topological
  censorship}, Phys. Lett. B \textbf{505} (2001), no.~1-4, 255--262.

\bibitem{gal:min}
Gregory~J. Galloway, \emph{Minimal surfaces, spatial topology and singularities
  in space-time}, J. Phys. A \textbf{16} (1983), no.~7, 1435--1439.

\bibitem{gal:cauchy}
\bysame, \emph{Some results on {C}auchy surface criteria in {L}orentzian
  geometry}, Illinois J. Math. \textbf{29} (1985), no.~1, 1--10.

\bibitem{galloway:nullsplit}
\bysame, \emph{Maximum principles for null hypersurfaces and null splitting
  theorems}, Ann. Henri Poincar\'e \textbf{1} (2000), no.~3, 543--567.

\bibitem{graham:lee}
C.~Robin Graham and John~M. Lee, \emph{Einstein metrics with prescribed
  conformal infinity on the ball}, Adv. Math. \textbf{87} (1991), no.~2,
  186--225.

\bibitem{HE}
S.~W. Hawking and G.~F.~R. Ellis, \emph{The large scale structure of
  space-time}, Cambridge University Press, London, 1973, Cambridge Monographs
  on Mathematical Physics, No. 1.

\bibitem{lawson}
H.~Blaine Lawson, Jr., \emph{Minimal varieties in real and complex geometry},
  Les Presses de l'Universit\'e de Montr\'eal, Montreal, Que., 1974,
  S\'eminaire de Math\'ematiques Sup\'erieures, No. 57 (\'Et\'e 1973).

\bibitem{mcinnes} B. McInnes, \emph{Exploring the similarities of the 
dS/CFT and AdS/CFT correspondences}, Nucl. Phys. B627 (2002) 311, 
\emph{The dS/CFT correspondence and the big smash}, hep-th/0112066, 
\emph{dS/CFT, censorship, instability of hyperbolic horizons, and
spacelike branes, hep-th/0205103.} 


\bibitem{ON}
Barrett O'Neill, \emph{Semi-{R}iemannian geometry}, Academic Press Inc.
  [Harcourt Brace Jovanovich Publishers], New York, 1983, With applications to
  relativity.

\bibitem{spivak:V}
Michael Spivak, \emph{A comprehensive introduction to differential geometry.
  {V}ol. {V}}, second ed., Publish or Perish Inc., Wilmington, Del., 1979.

\bibitem{strominger:ds/CFT:corr}
A.~Strominger, \emph{The {dS/CFT} correspondence}, hep-th/0106113, 2001.

\bibitem{strominger:ds/CFT:infl}
\bysame, \emph{Inflation and the {dS/CFT} correspondence}, hep-th/0110087,
  2001.

\bibitem{Wang}
X.~Wang, \emph{On conformally compact {Einstein} manifolds}, Math. Res. Letters
  \textbf{8} (2001), 1--18.

\bibitem{witten:AdS-holog}
E.~Witten, \emph{Anti de {S}itter space and holography}, Adv. Theor. Math.
  Phys. \textbf{2} (1998), no.~2, 253--291.

\bibitem{witten:talk}
\bysame, \emph{Talk at the {ITP} conference on {N}ew {D}imensions in {F}ield
  {T}heory and {S}tring {T}heory},
  http://online.itp.ucsb.edu/online/susy\_c99/witten/, November 1999.

\bibitem{witten:QG}
\bysame, \emph{Quantum gravity in de {Sitter} space}, hep-th/0106109, 2001.

\bibitem{witten:yau}
E. Witten and S.-T. Yau, \emph{Connectedness of the boundary in the
  {A}d{S}/{C}{F}{T} correspondence}, Adv. Theor. Math. Phys. \textbf{3} (1999),
  no.~6, 1635--1655 (2000).

\end{thebibliography}
\providecommand{\bysame}{\leavevmode\hbox to3em{\hrulefill}\thinspace}

\end{document}